%% file: main_0.tex
\documentclass[aps,twocolumn,amsmath,amssymb,nofootinbib,superscriptaddress,floatfix,reprint]{revtex4-2} %longbibliography,prl
\usepackage[latin1]{inputenc}
\usepackage{amsmath}
\usepackage{amsfonts}
\usepackage{amssymb}
\usepackage{lipsum}
\usepackage{bbold}
\usepackage{graphicx}
\usepackage{hyperref}
\usepackage{url}
\usepackage[usenames,dvipsnames]{color}
\usepackage{float}
\usepackage{multirow}
\usepackage{dcolumn}
\usepackage{soul}
\usepackage{physics}
\usepackage{braket}
\usepackage{subcaption}
\usepackage[range-units=single,separate-uncertainty]{siunitx}
\usepackage{placeins} %allows for floatbarrier, only used in supplement
\usepackage{verbatim} % for word count
\usepackage{layouts}
\usepackage{booktabs} %for tables
\usepackage[normalem]{ulem}     %sout, strikethrough in math mode
\usepackage{cancel} %cancel/strikethrough in math mode

\renewcommand{\arraystretch}{1.4} % increase the table cell height
\newcommand\thefont{\expandafter\string\the\font} % get information on font
\usepackage{centernot}
\newcommand{\fsl}[1]{{\centernot{#1}}}
\newcommand\xoutC{\bgroup \markoverwith{/\,}\ULon}

\newcommand\mxoutC{\bgroup \markoverwith{/}\ULon}

\usepackage{blindtext}
\newcommand{\pe}[1]{\textcolor{Red}{(P.E.: #1)}}

\newcommand{\yf}[1]{\textcolor{Purple}{(Y.F.: #1)}}

\begin{document}
%TC:ignore

\author{Philipp Eck}
 \affiliation{Institut f\"ur Theoretische Physik und Astrophysik and W\"urzburg-Dresden Cluster of Excellence ct.qmat, Universit\"at W\"urzburg, W\"urzburg, Germany}
 
\author{Yuan Fang}
 \affiliation{Department of Physics and Astronomy, Stony Brook University, Stony Brook, New York 11794, USA}

\author{Domenico Di Sante}
\affiliation{Department of Physics and Astronomy, Alma Mater Studiorum, University of Bologna, 40127 Bologna, Italy}
\affiliation{Center for Computational Quantum Physics, Flatiron Institute,162 5th Avenue, New York, New York 10010, USA}

\author{Giorgio Sangiovanni}
 \email{e-mail: sangiovanni@physik.uni-wuerzburg.de}
 \affiliation{Institut f\"ur Theoretische Physik und Astrophysik and W\"urzburg-Dresden Cluster of Excellence ct.qmat, Universit\"at W\"urzburg, W\"urzburg, Germany}
\author{Jennifer Cano}
 \email{e-mail: jennifer.cano@stonybrook.edu}
 \affiliation{Department of Physics and Astronomy, Stony Brook University, Stony Brook, New York 11794, USA}
\affiliation{Center for Computational Quantum Physics, Flatiron Institute,162 5th Avenue, New York, New York 10010, USA}

%%%%%%%%
%   Header
%%%%%%%%

%\title{Symmetry breaking stabilized HOTI phase on the triangular lattice\\
%Symmetry breaking stabilized topological phases on the triangular lattice\\
%From first order to higher order topology on the triangular lattice\\
%Recipe for higher order topology on the triangular lattice}
\title{Recipe for higher-order topology on the triangular lattice}

\date{\today}
\input{abstract_0}

\maketitle

%%%%%%%%%%%%%%%%%%%%%%%%%
%		Introduction
%%%%%%%%%%%%%%%%%%%%%%%%%
\input{introduction_0.tex}

\input{Z_2_sym_0.tex}

%%%%%%%%%%%%%%%%%%%%%%%%%
%   EBRS and dipole/quadrupole moment
%%%%%%%%%%%%%%%%%%%%%%%%%
\input{IRREP_polarization_0.tex}

%%%%%%%%%%%%%%%%%%%%%%%%%
%   finite size calculations 
%   of the model Hamiltonian
%%%%%%%%%%%%%%%%%%%%%%%%%
\input{finite_size_model_0.tex}

%%%%%%%%%%%%%%%%%%%%%%%%%
%   Ab initio finite size 
%   calculations
%%%%%%%%%%%%%%%%%%%%%%%%%
\input{finite_size_DFT_0.tex}

%%%%%%%%%%%%%%%%%%%%%%%%%%%
%  Conclusion
%%%%%%%%%%%%%%%%%%%%%%%%%%%
\input{conclusion_0.tex}

%%%%%%%%%%%%%%%%%%%%%%%%%%%
%  Methods section
%%%%%%%%%%%%%%%%%%%%%%%%%%%
%TC:ignore
%\input{methods_0.tex}

%{\noindent	\textbf{Data Availability} }
%\begin{scriptsize}
	%The data that support the plots within this paper and other findings of this study are available from the corresponding author upon reasonable request.
%\end{scriptsize}
%TC:endignorex

%%%%%%%%%%%%%%%%%%%%%%%%%%%
%  Acknowledgements
%%%%%%%%%%%%%%%%%%%%%%%%%%%

{\noindent
	\textbf{Acknowledgements}
}

P.E. und G.S. are grateful to Ralph Claessen for interesting discussions and thank the Flatiron Institute for the support and hospitality in the framework of a stimulating scientific cooperation.
J.C. and Y.F. acknowledge support from the National Science Foundation under Grant No. DMR-1942447.
P.E. and G.S. are grateful for funding support from the Deutsche Forschungsgemeinschaft (DFG, German Research Foundation) under Germany's Excellence Strategy through the W\"urzburg-Dresden Cluster of Excellence on Complexity and Topology in Quantum Matter ct.qmat (EXC 2147, Project ID 390858490) as well as through the Collaborative Research Center SFB 1170 ToCoTronics (Project ID 258499086). The research leading to these results has received funding from the European Union's Horizon 2020 research and innovation programme under the Marie Sk\l{}odowska-Curie Grant Agreement No. 897276. We gratefully acknowledge the Gauss Centre for Supercomputing e.V. (www.gauss-centre.eu) for funding this project by providing computing time on the GCS Supercomputer SuperMUC-NG at Leibniz Supercomputing Centre (www.lrz.de).
The Flatiron Institute is a division of the Simons Foundation.

%%%%%%%%%%%%%%%%%%%%%%%%%%%
%  Author contributions
%%%%%%%%%%%%%%%%%%%%%%%%%%%

%{\noindent	\textbf{Author contributions}}

%%%%%%%%%%%%%%%%%%%%%%%%%%%
%  Bibliography
%%%%%%%%%%%%%%%%%%%%%%%%%%%

%\bibliography{biblio_0.bib}
\input{main_0.bbl}
%%%%%%%%%%%%%%%%%%%%%%%%%%%
%  Supplement
%%%%%%%%%%%%%%%%%%%%%%%%%%%
\newpage
\section{Supplement}
\input{supplement/tb_model_0}
\input{supplement/DFT_supp}
\input{supplement/char_tab}
%\clearpage
%\section{Sections of the Supplement to be removed}
%\input{V0/supplement/do_not_show_supp}

% Don't count these!
%TC:ignore
%\quickwordcount{main}
%\quickcharcount{main}
%\detailtexcount{main}
%TC:endignore
\end{document}

%% file: abstract_0.tex
\begin{abstract}
    We present a recipe for an electronic 2D higher order topological insulator (HOTI) on the triangular lattice that can be realized in a large family of materials.
    The essential ingredient is mirror symmetry breaking, which allows for a finite quadrupole moment and trivial $\mathbb{Z}_2$ index.
    The competition between spin-orbit coupling and the symmetry breaking terms gives rise to four topologically distinct phases; the HOTI phase appears when symmetry breaking dominates, including in the absence of spin-orbit coupling.
    We identify triangular monolayer adsorbate systems on the (111) surface of zincblende/diamond type substrates as ideal material platforms and predict the HOTI phase for $X=$(Al,B,Ga) on SiC.
\end{abstract}

%% file: introduction_0.tex
%\section*{Introduction}
%\label{sec:Introduction}

\emph{Introduction}.---
A higher-order topological insulator (HOTI) is a new phase of matter that is gapped in its bulk and on its surfaces but exhibits gapless modes localized on hinges or corners where two surfaces meet \cite{benalcazar2017quantized,benalcazar2017electric,langbehn2017reflection,song2017d,schindler2018higher,schindler2018higher2}. 
Following the discovery of HOTIs, bismuth was immediately realized as a three-dimensional HOTI \cite{schindler2018higher2}.
In two dimensions (2D), 
%a HOTI is an example of an obstructed atomic limit \cite{bradlyn2017topological}.
%2D 
HOTIs were originally predicted in cold atoms \cite{benalcazar2017quantized} and have been realized in metamaterials \cite{serra2018observation,peterson2018quantized,imhof2018topolectrical,xue2019acoustic,ni2019observation,noh2018topological,fan2019elastic}.
However, an experimental demonstration of a 2D HOTI in an electronic system is still lacking.

In this manuscript, we present a tunable recipe for an electronic 2D HOTI that can be realized in a large class of hexagonal and trigonal material platforms.
The theory is built on an angular momentum $l=1$ (sub-) shell on the triangular lattice.
The essential new ingredient is symmetry breaking: 
%The essential ingredient is an angular momentum $l=1$ (sub-) shell on the triangular lattice with mirror- and inversion-symmetry breaking.
specifically, the absence of the horizontal reflection plane is necessary to open a hybridization gap, while the absence of the vertical reflection plane allows for a non-vanishing quadrupole moment.
Thus, mirror and inversion symmetry breaking is essential to realize the resulting HOTI phase: the phase is forbidden on the fully symmetric triangular lattice in this model.
In addition, the HOTI does not require spin-orbit coupling (SOC): when the symmetry breaking is small, SOC opens a trivial gap, while it plays no role when the symmmetry breaking dominates.
These features are in contrast to the famous Kane-Mele model \cite{kane2005z},
where infinitesimal SOC opens a topological gap and inversion symmetry breaking ultimately trivializes the quantum spin Hall insulator (QSHI).
%where inversion symmetry breaking ultimately trivializes the quantum spin Hall insulator (QSHI) without realizing a HOTI.
In fact, as we will show below, the HOTI phase cannot be achieved within the Kane-Mele model.

The main innovation of our work is to present a unified and realistic theory of HOTIs on the triangular lattice.
Our analysis of elementary band representations (EBRs) \cite{bradlyn2017topological,po2017symmetry,cano2018building} gives insight into the physical mechanism behind corner charge driven by symmetry breaking. It includes earlier predictions of HOTIs in inversion-breaking transition-metal dichalcogenides \cite{wang2019higher,zeng2021multiorbital,qian2022c} and is simpler than proposals requiring multiple atoms in the unit cell \cite{ezawa2018higher,liu2019two,sheng2019two,park2019higher,lee2020two,xue2021higher,costa2021discovery}.
%Further, our model is distinct from HOTIs that rely on SOC~\cite{costa2021discovery}.
Identifying the essential ingredients allows us to make material predictions based on symmetry criteria, which we verify by first principles calculations; one example is aluminum deposited on SiC.

%% file: Z_2_sym_0.tex
\begin{figure}
    \centering
    \includegraphics[width=\columnwidth]{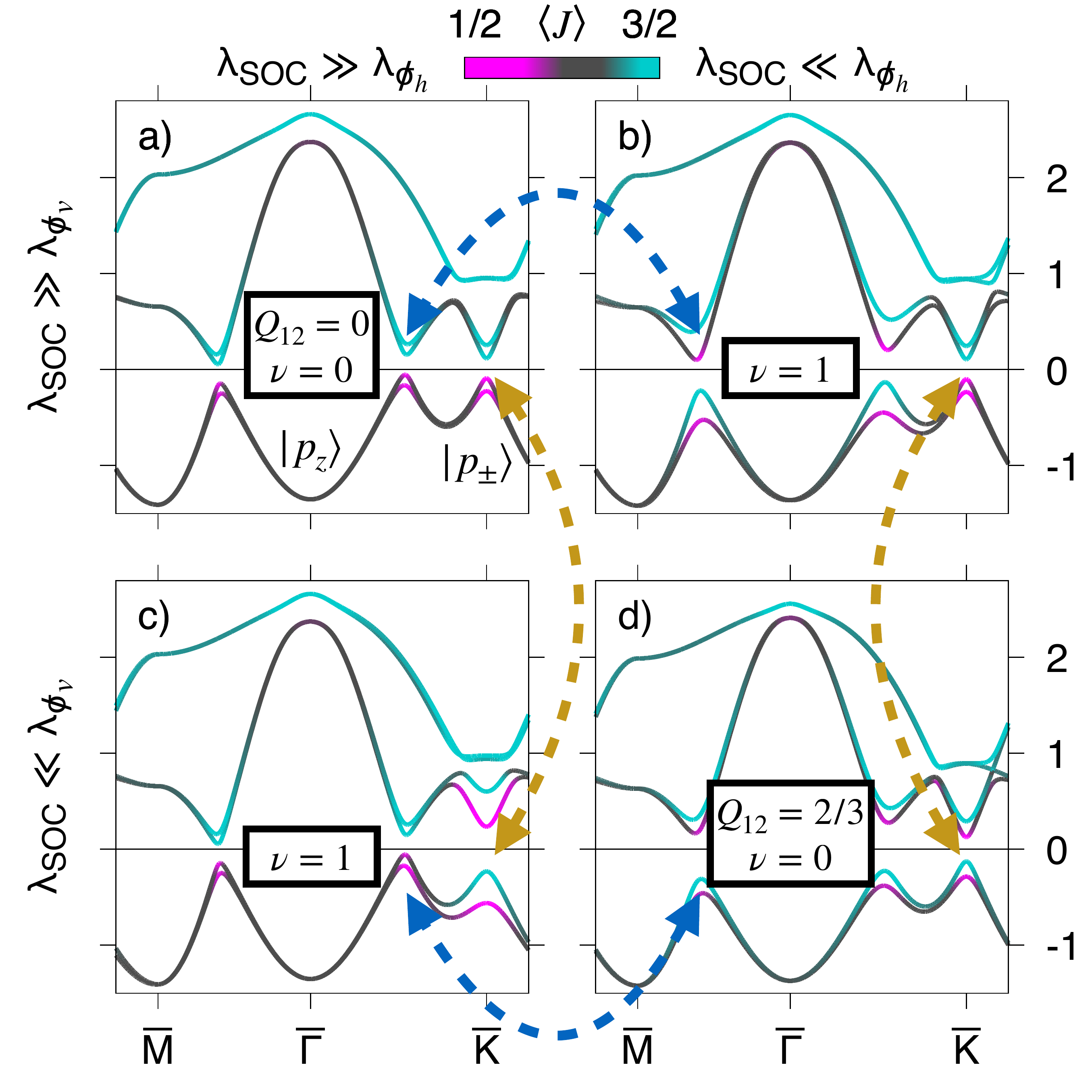}
    \caption{Band structures indicating the $\mathbb{Z}_2$ topological invariant $\nu$ and quadrupole moment $Q_{12}$ of the topologically distinct phases on the triangular lattice. The color code denotes the $\langle J \rangle$ character and the arrows indicate the relevant band inversion between neighboring phases. The labels in \textbf{a)} denote the dominant orbital character of the valence bands.}
    \label{fig:fig_Z2_bands}
\end{figure}

\emph{Topological phases driven by symmetry breaking}.---
We present a general model that describes $p$ orbitals, or, more generally, an $l=1$ angular momentum sub-shell, on the triangular lattice with tunable in-plane and out-of-plane mirror symmetry breaking terms and spin-orbit coupling.
By varying these parameters, the model realizes four phases, as depicted in Fig.~\ref{fig:fig_Z2_bands}.
Figure \ref{fig:fig_Z2_bands} also reveals the surprising property that the symmetry-breaking terms are indispensable to realizing non-trivial topology:
specifically, when local SOC dominates over all symmetry-breaking terms,
the ground state is topologically trivial, 
while in the limit of vanishing SOC, the HOTI phase is realized.
When only one symmetry-breaking term dominates over spin-orbit coupling, the system is in a $\mathbb{Z}_2$ QSHI phase.
The QSHI phase shown in Fig.~\ref{fig:fig_Z2_bands}b was recently realized in indenene, where symmetry breaking is provided by a SiC substrate
\cite{wang2016quantum,si2016large,chen2018large,bauernfeind2021design}.
%: as depicted in Fig.~\ref{fig:fig_Z2_bands}, the QSHI phase is realized only when the mirror or inversion symmetry breaking terms dominate over spin orbit coupling. 

%Here we elaborate on symmetry breaking induced topological phases of a generic angular momentum (sub-) shell basis with magnetic quantum number $m=\{0,-1,1\}$ and show that the phase diagram also realizes a HOTI.

The model is described by the Hamiltonian
\begin{align}
   \hat{H}=\hat{H}^{T}+\lambda_{SOC}\hat{H}^{SOC}+\lambda_{\fsl{\sigma}_h}\hat{H}^{\fsl{\sigma}_h}+\lambda_{\fsl{\sigma}_v}\hat{H}^{\fsl{\sigma}_v},
   \label{eq:H_lattice}
\end{align}
where each term is written explicitly in Supp.~\ref{sup:tb_model}.
The first term, $\hat{H}^{T}$, describes the symmetry allowed nearest-neighbor hoppings in the inversion symmetric triangular lattice layer group (LG) p6$/mmm$, generated by a six-fold rotation, three vertical reflection planes $\sigma_v$, three diagonal reflection planes $\sigma_d$, and one horizontal reflection plane $\sigma_h$. 
The second term, $\hat{H}^{SOC}$, is the local SOC interaction, which preserves the layer group symmetry and gaps the nodal line inside the BZ as well as the Dirac cones at the valley momenta; the gaps opened by SOC can be seen along
 $\overline{\Gamma\text{M}}$ and $\overline{\Gamma\text{K}}$ and at $\overline K/\overline K^\prime$. 
Each of the two remaining terms breaks inversion symmetry in addition to a reflection symmetry.
We use a strikeout notation to indicate the broken reflection symmetry. The third term, $\hat{H}^{\fsl{\sigma}_h}$, breaks $z\mapsto -z$,
which reduces the LG down to p6$mm$. 
It allows for hybridization between the states with magnetic quantum numbers $m=0$ and $m=\pm1$, i.e., it gaps the nodal line described above, which is formed when the $p_{\pm}$ bands cross the $p_z$ bands.
%Except at the high symmetry momenta, it allows for hybridization between the states with magnetic quantum numbers $m=0$ and $m=\pm1$, i.e., it opens a gap when the $p_{x/y}$ bands cross the $p_z$ bands \jc{does this happen along the same nodal line that is gapped by SOC?}\pe{Yes, the nodal line can bei either gapped by SOC or by the $\fsl{\sigma}_h$ term}. 
Finally, the last term, $\hat{H}^{\fsl{\sigma}_v}$, 
breaks vertical reflection ($\sigma_v$) and ($C_2$) rotation. The six-fold rotation ($C_6$) reduces to $C_3$, resulting in the LG p$\overline{6}m2$ (if $\lambda_{\fsl{\sigma}_h}=0$).
The absence of $\sigma_v$ lowers the little group at the valley momenta from $C_{3v}$ to $C_3$, splitting the two dimensional representation describing $p_+$ and $p_-$ orbitals into two one dimensional chiral representations ($-m,+m$) (for a more detailed discussion, see Supp.~\ref{sec:sup:vertical_reflection_sym},~\ref{sec:sup:character}). This term can be regarded as a non-local Semenoff mass term.

The competition between the inversion symmetry breaking terms and the atomic SOC determines the topological phase of the model. The four insulating phases are separated from each other by gap-closing phase transitions that exchange bands, as indicated by the arrows in Fig.~\ref{fig:fig_Z2_bands}.
Each gap reopening is accompanied by a band inversion that exchanges bands of predominately $J=1/2$ character with those of $J=3/2$ character, shown by the colors in Fig.~\ref{fig:fig_Z2_bands}.
Simultaneously, the band inversion changes the $\mathbb{Z}_2$-invariant, $\nu$, computed by tracking the Wilson loop eigenvalues \cite{soluyanov2011computing,yu2011equivalent}.
The results can be summarized as follows:
%The fundamental changes in the band structure, driven by the competition of SOC and symmetry breaking terms, is reflected in the $\langle J\rangle$-character. A summary can be found in Table~\ref{tab:phase_overview}. 
when SOC dominates (Fig.~\ref{fig:fig_Z2_bands}a), the valence(conduction) bands have the same value of $\langle J \rangle$ across the BZ. This indicates a $\nu =0$ topologically trivial insulator, where the valence(conduction) bands transform as an atomic limit with $J=1/2(J=3/2)$.
We dub this phase an ``SOC insulator''.
By breaking either reflection symmetry, $\sigma_h$ or $\sigma_v$, a $\nu = 1$ QSHI phase can be reached: 
in the former case, the hybridization between the $p_z$ and the in-plane orbitals dominates over the SOC term along the nodal line,
%\pe{except at the high symmetry points \textit{or} at valley momenta and at $\Gamma$ (\textit {important for the $C_3$ eigenvalue discussion in the next section})}
%\jc{not sure why you wrote ``except the high symmetry points''. I thought the band inversion at $K$ was important}\pe{right, here I wanted to stress that the $\fsl{\sigma}_h$ term dominates over SOC inside the BZ, but at K the $\fsl{\sigma}_h$ term vanishes the valley momenta will be gapped by SOC. In other words, the $\fsl{\sigma}_h$ doesn't render the little group of K abelian.}
stabilizing an indenene-like QSHI phase (Fig.~\ref{fig:fig_Z2_bands}b) \cite{bauernfeind2021design}.
The other QSHI phase is characterized by a strong local orbital angular momentum polarization at the valley momenta, which gaps the in-plane Dirac bands (``{$\fsl{\sigma}_v$} QSHI'', Fig.~\ref{fig:fig_Z2_bands}c).
Finally, if both symmetry breaking terms dominate over SOC, or if SOC is absent, the $\mathbb{Z}_2$-index vanishes (Fig.~\ref{fig:fig_Z2_bands}d) again. However, the resulting insulator phase is not trivial: as we will show momentarily, it has a nontrivial polarization and filling anomaly, indicating that it is a HOTI and exhibits corner charge on a finite-sized lattice. 

%% file: IRREP_polarization_0.tex
%\section{Symmetry indicators and polarization}
%\label{sec:Irreps_pol}
\begin{comment}
\begin{itemize}
    \item Derive from the symmetry analysis the Irreps for the four $Z_2$ phases
    \item the competition of SOC and the ISB term in the periodic direction determine the Irreps at the valley momenta, Irreps are unaffected by the mirror symmetry breaking
    \item Based on the Irreps, give the dipole and quadrupole moment
    \item Conclusion: Irreps promotes dipole and quadrupole moment, filling anomaly predicted for the "boring symmetry broken insulator?" phase.
\end{itemize}
\end{comment}

\renewcommand{\arraystretch}{1.3}
\setlength{\tabcolsep}{5pt}

\begin{table*}[]
\begin{tabular}{cccccccc}
\multicolumn{1}{l}{}                & \multicolumn{1}{l}{} & \multicolumn{1}{l}{} & \multicolumn{1}{l}{}                            & \multicolumn{1}{l}{}                            & \multicolumn{1}{l}{}          & \multicolumn{1}{l}{}                   & \multicolumn{1}{l}{}   \\ \hline
Phase                               & Layer Group          & $\nu$                & SOC vs $\fsl{\sigma}_h$                         & SOC vs $\fsl{\sigma}_v$                         & $\xi(C_3)$ at  $\overline{K}$ & $\mathbf{P}=(P_1,P_2)$                 & $Q_{12}$               \\ \hline
SOC insulator                       & p6/$mmm$             & 0                    & $\lambda_\text{SOC}\gg\lambda_{\fsl{\sigma}_h}$ & $\lambda_\text{SOC}\gg\lambda_{\fsl{\sigma}_v}$ & $\{e^{+i\pi/3},e^{-i\pi/3}\}$ & $(0,0)~\text{mod}~2$                   & $0~\text{mod}~1$       \\
Indenene-like $\fsl{\sigma}_h$ QSHI & p6$mm$               & 1                    & $\lambda_\text{SOC}\ll\lambda_{\fsl{\sigma}_h}$ & $\lambda_\text{SOC}\gg\lambda_{\fsl{\sigma}_v}$ & $\{e^{+i\pi/3},e^{-i\pi/3}\}$ & -                                      & -                      \\
$\fsl{\sigma}_v$ QSHI               & p$\overline{6}m2$    & 1                    & $\lambda_\text{SOC}\gg\lambda_{\fsl{\sigma}_h}$ & $\lambda_\text{SOC}\ll\lambda_{\fsl{\sigma}_v}$ & $\{e^{\pm i\pi/3},-1\}$       & -                                      & -                      \\
Triangular HOTI                     & p$3m1$               & 0                    & $\lambda_\text{SOC}\ll\lambda_{\fsl{\sigma}_h}$ & $\lambda_\text{SOC}\ll\lambda_{\fsl{\sigma}_v}$ & $\{e^{\pm i\pi/3},-1\}$       & $(\mp\frac23,\mp\frac23)~\text{mod}~2$ & $\frac23~\text{mod}~1$ \\ \hline
%\multicolumn{1}{l}{}                & \multicolumn{1}{l}{} & \multicolumn{1}{l}{} & \multicolumn{1}{l}{}                            & \multicolumn{1}{l}{}                            & \multicolumn{1}{l}{}          & \multicolumn{1}{l}{}                   & \multicolumn{1}{l}{}  
\end{tabular}
\caption{$C_3$ rotation eigenvalues and dipole/quadrupole moments of the insulating phases of Eq.~(\ref{eq:H_lattice}). For each phase, the layer group indicated is the highest symmetry group that satisfies the inequalities in columns four and five. The electric multipoles in the $\nu=1$ phases are ill defined. The little groups, irreps and corresponding character tables of momenta $\overline \Gamma$ and $\overline K$ are shown in the Supp.~\ref{sec:sup:character}. 
}
\label{tab:phase_overview}
\end{table*}

\emph{Symmetry indicators and polarization}.---
The symmetry and topology of each phase is summarized in Table~\ref{tab:phase_overview}.
The strong topological invariants of the two $\nu=1$ phases are not symmetry-indicated due to the lack of inversion symmetry. However, the electric polarization and quadrupole moments of the HOTI and SOC insulating phases with $\nu= 0$ can be diagnosed by symmetry indicators~\cite{fang2021filling,fang2021classification,takahashi2021general,watanabe2020corner} constructed from the EBRs \cite{bradlyn2017topological,po2017symmetry,cano2018building}.

To compute the symmetry indicators,
we define lattice vectors $\mathbf a_1=(1,0)$, $\mathbf a_2=(1/2,\sqrt 3/2)$ and reciprocal lattice vectors $\mathbf b_1=2\pi(1,-1/\sqrt{3})$, $\mathbf b_2=(0,4\pi/\sqrt{3})$. The polarization vector with components in the directions of the two primitive lattice vectors is defined by $\mathbf P=(P_1,P_2)=-\langle (r_1,r_2) \rangle$, where $r_{1,2}$ are the relative coordinates of the point $\mathbf{r}=r_1\mathbf{a}_1 + r_2 \mathbf{a}_2$.
The quadrupole moment is given by $Q_{12}=-\langle r_1r_2+\frac14 (r_1^2+r_2^2) \rangle$ for a three-fold rotation symmetry \cite{watanabe2020corner}. 
%$Q_{12}=\langle r_1r_2-\cos\theta\frac{r_1^2+r_2^2}{2} \rangle$ where $\theta=2\pi/3$ for a three-fold rotation symmetry \cite{watanabe2020corner}. 
The symmetry indicators for polarization and quadrupole moment are \cite{fang2021filling,takahashi2021general,watanabe2020corner,fang2021classification}
\begin{align}
    P_1=P_2&= -\frac{2}{3}\left([\#e^{i\pi/3}]-[\#e^{-i\pi/3}]\right) \mod 2 \label{eq:dipole_mom}\\
    Q_{12}&= -\frac{2}{3}\left([\#e^{i\pi/3}]+[\#e^{-i\pi/3}]\right) \mod 1  \label{eq:quadru_mom}
\end{align}
where $[\# \xi]$ is the number of valence bands with $C_3$ eigenvalue $\xi=e^{i\frac{2\pi}{3}j_z}$ at $\overline\Gamma=\mathbf 0$ subtracted from the number of valence bands with $C_3$ eigenvalue $\xi$ at $\overline K=\frac23 \mathbf b_1+\frac13 \mathbf b_2$.

In all four phases the valence bands at $\overline\Gamma$ are always $p_z$-type with total magnetic quantum numbers $j_z=\{-1/2,+1/2\}$ (see Fig.~\ref{fig:fig_Z2_bands} a). Consequently, only the rotation eigenvalues at $\overline K$ can change the electronic polarization or quadrupole moment: the competition between $\hat H^{\text{SOC}}$ vs $\hat H^{\fsl{\sigma}_v}$ acting on the $p_\pm$ subspace
results in predominantly $j_z=\{-1/2,+1/2\}$ character in the valence bands when $\hat H^{\text{SOC}}$ dominates and $j_z=\{\pm 1/2,\pm 3/2\}$ character when $\hat H^{\fsl{\sigma}_v}$ dominates, where $\pm$ is determined by $\text{sign}(\lambda_v)$. %\pe{This depends on the sign of $\lambda_v$, we could take only the positive value, to have a consistent description of the $\mathbf{Q,P}$ and the Wyckoff position of the EBR in the HOTI phase.}.
%\yf{What about keeping $\pm$ both here and in the table by using $j_z=\text{sign}(\lambda_v) \{ 1/2, 3/2\}$?}
%following the notation of Refs.~\cite{bradley2010mathematical,bradlyn2017topological,vergniory2017graph,elcoro2017double} \pe{The references should be updated, do we need them here, if we don't give the references?}. \yf{You're right, we don't need these refs here.}
Applying Eqs.~(\ref{eq:dipole_mom}) and (\ref{eq:quadru_mom}), we find the following dipole and quadrupole moments for the two $\nu=0$ phases: the SOC insulator has $\mathbf P=0$, $Q_{12}=0$, while the triangular HOTI has $\mathbf P=\mp (2/3,2/3)\mod 2$, $Q_{12}= 2/3\mod 1$.
%\jc{Are the $\pm$ signs reversed here for $\mathbf{P}$ and $\mathbf{Q}$? For the $+$ sign, I thought $[\#e^{i\pi/3}]=0$, $[\#e^{-i\pi/3}]=-1$. Or maybe the text is backwards: it says $K-\Gamma$, maybe it shoould be $\Gamma-K$?}
These results are shown in Table~\ref{tab:phase_overview}. 
The non-zero quadrupole moment for the triangular HOTI phase implies the existence of corner localized states, which we study in the next section. \par

%We now phrase our results in terms of
Our results can be rephrased in terms of
EBRs~\cite{bradlyn2017topological,po2017symmetry,cano2018building}:
%From the irreducible representations (irreps) of the occupied bands, we find a characteristic shift in the localization of the EBRs~\cite{bradlyn2017topological,cano2018building} corresponding to the change in the electric polarization. 
the valence bands of the SOC insulator transform as an EBR induced from the irreducible representation (irrep) $\bar{E}_{1u}$ of the site-symmetry group at the $1a=(0,0)$ position, while the HOTI with  $\text{sign}(\lambda_v)=+1$ %\jc{do we need to add ``with  $\text{sign}(\lambda_v)=+1$''}
transforms as an EBR induced from the irrep $\bar E_1$ of the site-symmetry group of the $1b=(1/3,1/3)$ position. 
(The irrep notation follows Ref.~\cite{bradley2010mathematical}. The irreps corresponding to the valence bands in each phase are listed in the Supp.~\ref{sec:sup:character}). 

This change in EBRs indicates the transition to an obstructed atomic limit as the Wannier center shifts from $1a$ to $1b$, corresponding to the electronic charge center detaching from the lattice sites in the HOTI phase to create the nonzero polarization and quadrupole moment. 
Similarly, for $\text{sign}(\lambda_v)=-1$ the Wannier center shifts from $1a$ to $1c=(2/3,2/3)$, creating a nonzero polarization and quadrupole moment of the opposite sign.
Breaking the vertical mirror planes $\sigma_v$ is imperative to realize this phase: since $\sigma_v$ maps $1b=(1/3,1/3)$ onto $1c=(2/3,2/3)$, its presence forbids a Wannier center on $1b$ without a partner on $1c$ and vice versa (see Supp.~\ref{sec:sup:vertical_reflection_sym} for a more detailed discussion).

Note that such a HOTI phase cannot exist in the Kane-Mele model: a $\nu=0$ insulating ground state can only be reached by breaking inversion symmetry to gap the Dirac fermions~\cite{kane2005z}. In this phase, the Wannier functions are localized on one of the two atomic sublattices; consequently, the system lacks a finite dipole and quadrupole moment.

%% file: finite_size_model_0.tex
%\section{HOTI edge and corner charge}
%\label{sec:finite_size_model}

\begin{figure}[b]
    \centering
    \includegraphics[width=\columnwidth]{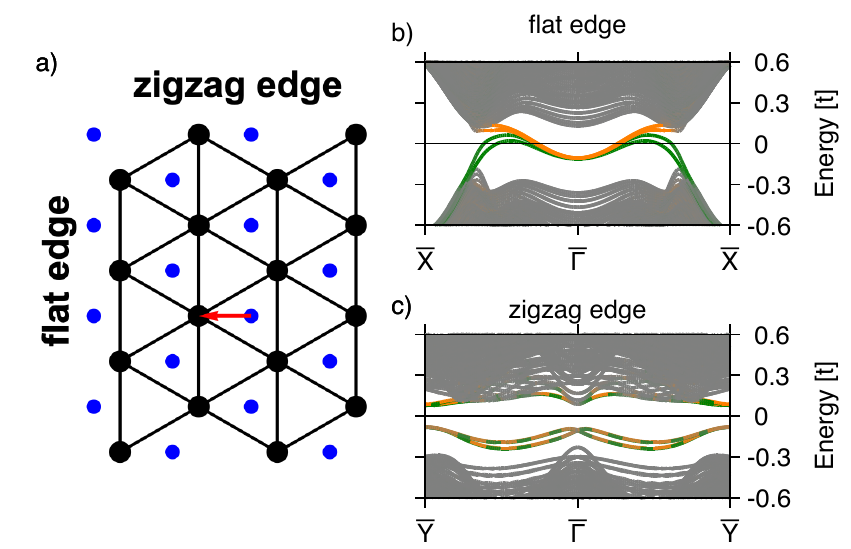}
    \caption{Polarization and slab calculations for the HOTI phase with SOC.
    a) The bulk dipole moment (red vector), resulting from Wannier centers located at the $1b$ Wyckoff position (blue dots) is perpendicular to the flat edge and parallel to the zigzag edge.
    b,c) Slab band structure and edge character (orange-green color code) for the two slab terminations. 
    The polarization parallel (perpendicular) to the edge in the zigzag (flat) geometry yields insulating (metallic) edge states.
    }
    \label{fig:model_slab}
\end{figure}

\emph{HOTI edge and corner charge}.---
The electronic dipole moment in the HOTI phase has important consequences for finite size geometries. As shown in Fig.~\ref{fig:model_slab}a, the triangular lattice has two canonical edge terminations: the zigzag and the flat edge.
The bulk polarization $\mathbf{P}$, arising from Wannier centers located at $1b$ (blue dots) in Fig.~\ref{fig:model_slab}a, is parallel to the zigzag edge and normal to the flat edge; the latter favors metallic edge states \cite{zeng2021multiorbital,bollinger2001one}. 
For the model, the edge states of the flat termination are non-degenerate and possess a linear band crossing at $\overline{\Gamma}$, as shown in Fig.~\ref{fig:model_slab}b (the touching is quadratic in the limit of vanishing SOC.) 
%\pe{quadratic wo SOC}. 
In contrast, the zigzag geometry has degenerate insulating edge states, shown in Fig.~\ref{fig:model_slab}c. 
% Remove the following, as we don't show the SOC insulator in the supplement.
%For the SOC insulator, which has vanishing bulk polarization $\mathbf{P}=0$, we find insulating and degenerate edge states for both edge terminations (see supplemental Fig.~\ref{fig:supp_slab_soc_ins}).

\begin{figure}
    \centering
    \includegraphics[width=\columnwidth]{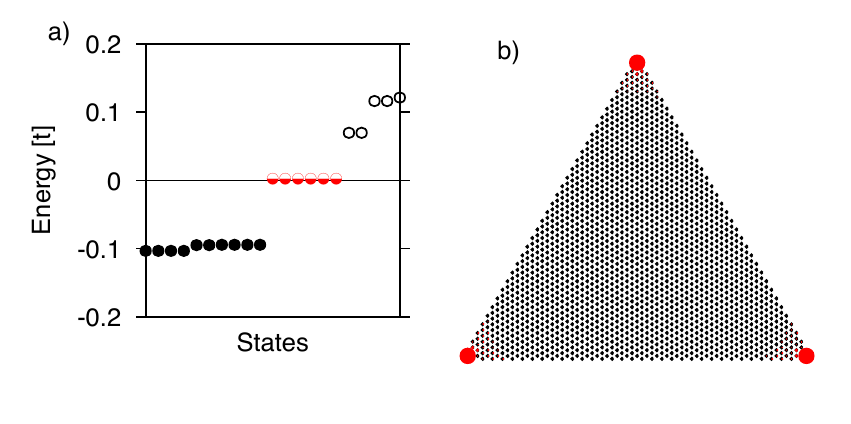}
    \caption{Triangular flake spectrum and charge localization for the HOTI. At charge neutrality, there are exactly two electrons to fill the six mid-gap states shown in red in a), in agreement with the filling anomaly $\eta = 2$.
    The point size in b) shows the wave function localization of the red mid-gap states in a).
    }
    \label{fig:model_flakes}
\end{figure}

To isolate the fractionally filled corner states living in the bulk and edge gaps, we consider triangular flakes with the insulating zigzag termination.
In the HOTI phase at charge neutrality, we find six degenerate exponentially corner-localized states that are one-third occupied at an energy within the bulk and edge gaps, as shown in Fig.~\ref{fig:model_flakes}a and in agreement with the corner charge of $Q_{12} = 2/3$ computed in the previous section.
That there are two electrons to occupy the six mid-gap states at charge neutrality is referred to as the ``filling anomaly'', $\eta= 3Q_{12} = 2$, where the factor of three corresponds to the three corners of the triangular flake~\cite{fang2021filling,benalcazar2019quantization,schindler2019fractional}.
%\pe{Isn't it $\#$corners times $Q_{12}$?})
%\jc{But we have three corners, so this is correct, yes? Or should we add, ``where the factor of three comes from the triangular geometry.''}
%\pe{I think, a reader not being an expert in the filling anomaly would have expected after reading the previous sentence, that $\nu=\#\text{degeneracy of edge states}Q_{12}$. Probably, we could add \textit{which arises from the finite quadrupole moment at the three corners}. On the other hand, we give references and it will be clear to experts, I'm also fine with the current version.}
%We don't show the localization in the supplement, mention it in the paragraph before.
%These symmetry-protected states are exponentially localized on the corners of the triangular flake, as shown in Fig.~\ref{fig:model_flakes}b and supplemental Fig.~\ref{fig:supp_model_edge_loc}. 

For the flat-edge termination with finite edge polarization and metallic edge states (Fig.~\ref{fig:model_slab}a), fractionally filled corner states can be only stabilized if the edge charge is compensated \cite{watanabe2020corner,fang2021filling,zeng2021multiorbital,qian2022c}.

%0D triangular flakes

%% file: finite_size_DFT_0.tex
%\section{Finite size ab initio calculations -- substrate induced HOTI phase}
%\section{Material realization}
%\label{sec:finite_size_DFT}

\emph{Material realization}.---
Having established the existence of the HOTI phase in our minimal triangular model, we propose a general material realization concept: triangular adsorbate systems on the high symmetry sites of the (111) surface of zinc-blende/diamond-type substrates. This substrate provides three important ingredients: 1) structural stabilization of a triangular adsorbate monolayer; 2) symmetry breaking across the horizontal mirror plane to open a hybridization gap ($\hat{H}^{\fsl{\sigma}_h}$); and 3) symmetry breaking across the vertical mirror planes ($\hat{H}^{\fsl{\sigma}_h}$) to induce the bulk quadrupole moment.
%\pe{Can we argue, that the quadrupole moment (actually hexapole moment) must vanish if we have $C_6$, but $C_3$ or $\overline{6}$ allows for a finite value? $\mathbb{Z}_2\times\mathbb{Z}_2$ HOTI, bismuth?}

\begin{figure}[b]
    \centering
    \includegraphics{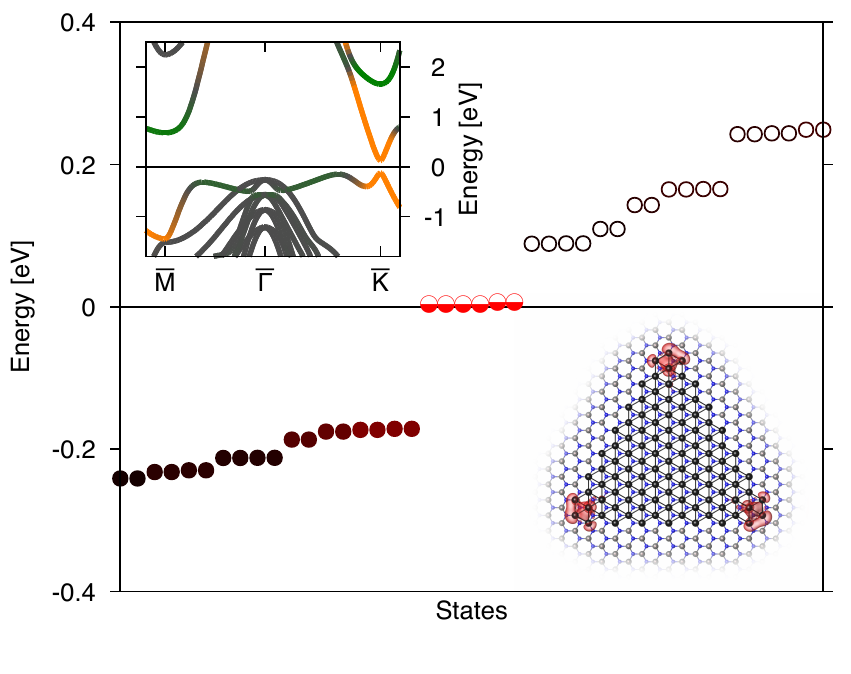}
    \caption{
    The energy spectrum of a finite-size triangular flake of Al on SiC, truncated as shown in the lower inset.
    The red color code denotes the corner character of the state: the six degenerate mid-gap states are completely localized on the corners.
    Upper inset: bulk band structure of Al on SiC; color code denotes the Al $p_z$ (green) and Al $p_\pm$ (orange) character.
    Lower inset: unit cell geometry and charge density of corner states: the $C_6$ symmetry of the Al (black) site on top of the Si atom (gray) is reduced to $C_3$ by the first C layer (blue).
    }
    \label{fig:DFT}
\end{figure}

We propose a monolayer of light Group 3 %\pe{Capital ''G'' in group?} \jc{Yes, Group is capitalized here.}
elements (B, Al, Ga) on SiC and verify our prediction with an \textit{ab initio} DFT study. For the T1 adsorption site of the Si-terminated surface, the adatom is located on top of the surface Si atom, while the C atom of the first SiC layer reduces the rotational symmetry of the triangular site from $C_6$ down to $C_3$ as shown in the inset of Fig.~\ref{fig:DFT}. In the case of Al,  in-plane and out-of plane reflection symmetry breaking (LG p$3m1$) dominates over SOC and results in an insulating bulk band structure with the $p_z$-type $\overline{\Gamma}_6(2)$ irrep and a $p_{\pm}$-type $\overline{K}_4(1)\oplus\overline{K}_6(1)$ irrep ($j_z=\{3/2,1/2\})$ in the valence bands (see also Tab.~\ref{tab:littlegroup6} in the supplemental material), identical to the triangular HOTI phase, as shown in Table~\ref{tab:phase_overview}.
Consequently, this phase has a quadrupole moment $Q_{12}=2/3~\text{mod}~1$ and a corresponding corner charge.

Varying the Group 3 elements, our \textit{ab initio} calculations reveal a valley momenta gap of $\Delta_\text{B}=0.49\,\text{eV}$, $\Delta_\text{Al}=0.24\,\text{eV}$ and $\Delta_\text{Ga}=0.36\,\text{eV}$. As shown in Fig.~\ref{fig:supp_HOTI_bulk}, only Al exhibits a direct band gap at the valley momenta; the global indirect band gap is $0.27\,\text{eV}$ and $0.18\,\text{eV}$ for B and Ga, respectively. 

We verify the symmetry indicated prediction of corner charge by a first principles calculation on a finite size lattice for Al on SiC. The insulating band structure for the zigzag termination is shown in the inset to Fig.~\ref{fig:DFT} (see also Fig.~\ref{fig:supp_DFT_ribbon} in the supplemental material). 
The calculation reveals six degenerate states in the bulk band gap, which are filled with two electrons at charge neutrality.
The charge density of these states are shown in the lower inset to Fig.~\ref{fig:DFT}, which are tightly localized to the corners. Furthermore, they display an almost perfect symmetry with respect to two of the three vertical mirror reflection planes of the bulk, even though these symmetries are broken at the edges and corners of the flake.

%% file: conclusion_0.tex
%\section*{Conclusion}
%\label{sec:Conclusion}

%\begin{itemize}
%    \item general model proposal: triangular lattice, $l$-sub-shell + ISB + MIR $\rightarrow$ can stabilize triangular HOTI
%    \item Highlight generality of the approach: huge abundancy of zincblende/diamond structure substrates (Si,C,GaAs,InSb...) and variety of potential adsorbates may path the way to further triangular HOTIs
%    \item bottleneck: besides experimental realization, insulating bulk/slab geometries
%\end{itemize}

\emph{Conclusion}.---
We have proposed a recipe for electronic HOTIs in materials where the low-energy bands are comprised of an $l=1$ angular momentum subshell. 
The essential ingredient is inversion- and reflection-symmetry breaking: on the symmetric triangular lattice, the HOTI phase is forbidden.
%The HOTI phase is driven by inversion- and reflection-symmetry breaking and does not rely on the presence of SOC.
We identified the HOTI phase using symmetry indicators and by an explicit calculation of the spectrum on a finite-sized triangular sample. 
%\pe{Further we highlighted the importance of vertical reflection symmetry breaking as it determines the rotation eigenvalues and allows for the formation of an obstructed atomic limit.} %% JC: I incorporated this as the second sentence of this paragraph, feel free to modify.

Our approach is very general and may be realized in many compounds by depositing adatoms onto the three-fold symmetric (111) surface of a zincblende/diamond substrate.
We identified by first-principles calculations the $\mathbb Z_2$-trivial analogues of the recently synthesized QSHI indenene~\cite{bauernfeind2021design}, namely B, Ga and Al on SiC, as potential candidates and showed explicitly for the case of Al a full finite-size study: it is bulk insulating and has gapped edges and localized corner charge on a finite-sized triangular flake.
Given the abundance of zincblende/diamond substrates (Si, C, GaAs, and InSb, for example), and a variety of potential adsorbates, we expect many other material combinations will also realize the HOTI phase.
Thus, our work paves the way to an experimental demonstration of a 2D electronic HOTI. 
A systematic \textit{ab initio} study of the material combinations to determine which are bulk insulators will be essential to future work.
%\jc{Such a study could also be extended to atoms with $d$- and $f$-orbitals to realize ``heavy'' HOTIs with sizeable electron-electron interactions and SOC.}
Upon extension to atoms with $d$- and $f$-orbitals, we expect ``heavy'' HOTIs with sizeable electron-electron interactions and SOC.
%\pe{How about heavy compounds, $d,f$-sub-shells with a sizable electron-electron and SOC interaction. Model relies on fundamental symmetries in lattices, applicable to cold atoms, photonics and phononics?
%@ Jen and Yuan, do you see strict limitations of our model to 2D? How about ``quasi 2D'' 3D layered heterostructures?}

%% file: main_0.bbl

%% file: supplement/tb_model_0.tex
\subsection{Tight-Binding Model}
\label{sup:tb_model}

Here we describe the tight binding Hamiltonian of a $p$-shell in the $\{p_x,p_y,p_z\}$-basis on a triangular lattice as shown in Fig.~\ref{fig:fig_Z2_bands} with the Bravais vectors $\mathbf{a}_1=(1,0)$ and $\mathbf{a}_2=(0.5,\sqrt{3}/2)$.

\subsubsection{Triangular lattice hopping Hamiltonian}

The transition matrix elements $H_{ij}^T$ allowed by the symmetries of LG p$6/mmm$ can be obtained by following the approach of Slater and Koster \cite{slater1954simplified}. They are given for an orbital $p_j$ located in the home unit cell ($\mathbf{0}$) to an orbital $p_i$ at site $\mathbf R$:
\begin{align}
	H_{ii}^T(\mathbf{R})=\langle p_i(\mathbf{0})| \hat H^T | p_i(\mathbf{R})\rangle &= n_i^2V_i^\sigma+(1- n_i^2)V_i^\pi,
	\label{eq:pipi_supp}\\
	H_{ij}^T(\mathbf{R})=\langle p_i(\mathbf{0})| \hat H^T | p_j(\mathbf{R})\rangle &= n_i n_j (V_{i,j}^\sigma-V_{i,j}^\pi), \label{eq:pipj_supp}
\end{align}
with $i=x,y,z$ and $i\neq j$. The coefficients $n_i$ incorporate the in-plane orientation ($n_x=\cos(\phi)\sin(\theta), n_y=\sin(\phi)\sin(\theta)$ and $n_z=\cos(\theta)$) with the azimuthal angle $\phi(\mathbf{R})$ and polar angle $\theta(\mathbf{R})$. The transfer integral values $V^\sigma$ and $V^\pi$ in the $p_{xy}$ subspace, the $p_z$ transfer integral $V_{p_z}^\pi$ and the on-site energy shift of the $p_z$ orbital $E_z$ are given in Table~\ref{tab:tb_param}. 
The strength of the SOC interaction and the symmetry breaking terms of the relevant layer groups are listed in Table~\ref{tab:tb_param_LG}.
All tight-binding parameters have been chosen such that an insulating ground state in the corresponding phase is stabilized. The overall band character reflects qualitatively the low energy-band structure of the Group 3 elements on SiC, with a $p_z$ and $p_{xy}$ valence band character at $\overline \Gamma$ and $\overline K$, respectively.

\subsubsection{Atomic SOC}
We consider full $p$-shell atomic spin orbit coupling, which is given in the $\{p_x,p_y,p_z\}$-basis by:
\begin{align}
    \hat{H}^{SOC}=&\lambda_{SOC} \hat{L} \otimes \hat{S} \\
                  =& \frac{\lambda_{SOC}}{2}\left( \begin{array}{rrr}
                0\hphantom{_x} & -i\sigma_z & i\sigma_y \\
                i\sigma_z & 0\hphantom{_x} & -i\sigma_x  \\
                -i\sigma_y & i\sigma_x & 0\hphantom{_x}   
\end{array}\right).
\end{align}
Its matrix elements can be obtained by explicitly calculating the components of the orbital angular momentum and spin operators.

\subsubsection{$\sigma_h$-Symmetry Breaking}
The presence of vertical reflection symmetry prohibits the hybridization between the in-plane and out-of plane orbitals.
When the symmetry is broken, the Slater-Koster integrals in Eqs.~\ref{eq:pipi_supp} and \ref{eq:pipj_supp} become non-zero because the out-of plane coordinates of the $p_z$ and the in-plane orbitals differ, i.e., the polar angle $\theta \neq\pi/2$. The effective transfer elements read:
\begin{align}
	H_{iz}^{\fsl{\sigma}_h}(\mathbf{R}) = \langle p_i(\mathbf{0})| \hat H^{\fsl \sigma_h} | p_z(\mathbf{R})\rangle &=+ n_i \lambda_{\fsl \sigma_h}, \label{eq:pipj_mir1} \\
	H_{zi}^{\fsl{\sigma}_h}(\mathbf{R})=\langle p_z(\mathbf{0})| \hat H^{\fsl \sigma_h} | p_i(\mathbf{R})\rangle &= -n_i \lambda_{\fsl \sigma_h}, \label{eq:pipj_mir2}
\end{align}
with $\lambda_{\fsl \sigma_h}=n_z(\theta) (V_{xy,z}^\sigma-V_{xy,z}^\pi)$.

\subsubsection{$\sigma_v$-Symmetry Breaking}

To break $\sigma_v$ while preserving $\sigma_d$ requires breaking $C_{2z}$. 
%\jc{The $\sigma_v$-symmetry breaking term also breaks $C_2$-rotation symmetry. -- Philipp is there a way to explain this better by describing the physical mechanism you have in mind? My point is that there are multiple ways to break $\sigma_v$ and we should specify why we are choosing this particular way, which breaks $C_{2z}$ and $\sigma_v$ together.}
%\pe{I think, to break $C_{2z}$ one of the vertical mirrors ($\sigma_d$ and $\sigma_v$ must be broken, since they have reflection lines, which are normal to each other $C_{2z}=\sigma_{d,i} \sigma_{v,j}$, but only the absence of $\sigma_v$ results in our model to inequivalent Wyckoff positions on the $1b$ and $1c$ site and allows for a gapping at the valley momenta. I think, this is the only way, to break $\sigma_v$, as long as we keep the Bravais vectors fixed or do I miss something?}
%The presence of $\sigma_v$ and $C_2$ symmetry guarantees for symmetric hoppings under an inversion of the hopping direction. In the absence of $C_2$, they can become antisymmetric but must respect the remaining $C_3$ symmetry. 
The absence of $C_{2z}$ symmetry allows for the hopping terms to become asymmetric when the hopping direction is reversed.
Since they must still respect the three-fold rotation symmetry $C_{3z}$,
such an interaction can be described by the following transfer matrix elements:
\begin{align}
     H_{yx}^{\fsl \sigma_v}(\mathbf{R})=\langle p_y(\mathbf{0})| \hat H^{\fsl \sigma_v} | p_x(\mathbf{R})\rangle &= +\lambda_{\fsl \sigma_v} \cos(3\phi), \label{eq:isb_pyx}\\
    H_{yx}^{\fsl \sigma_v}(\mathbf{R})=\langle p_x(\mathbf{0})| \hat H^{\fsl \sigma_v} | p_y(\mathbf{R})\rangle &= -\lambda_{\fsl \sigma_v} \cos(3\phi), \label{eq:isb_pxy}
\end{align}
where $\phi(\mathbf{R})$ is the azimuthal angle. The opposite sign in Eqs.~\ref{eq:isb_pyx} and \ref{eq:isb_pxy} is a consequence of the broken $C_2$ symmetry.

\renewcommand{\arraystretch}{1.5}
\setlength{\tabcolsep}{12pt}

\begin{table}
    \centering
    \begin{tabular}{ c c c c }
    \hline
    $E_z$ & $V^\sigma$ & $V^\pi$ & $V_{p_z}^\pi$ \\ \hline
    -0.7  & 0.7        & -0.15   & -0.25  \\ \hline      
    \end{tabular}
    \caption{Tight-binding parameters of the model Hamiltonian $H^T$ in units of $t$. 
    }
    \label{tab:tb_param}
\end{table}

\begin{table}
    \centering
    \begin{tabular}{ c c c c }
    \hline
    LG               & $\lambda_\text{SOC}$ & $\lambda_{\fsl{\sigma}_h}$ & $\lambda_{\fsl{\sigma}_v}$ \\ \hline
    p$6/mmm$         & 0.15                 & 0.1/6                      & 0.04/3                     \\
    p$6mm$           & 0.15                 & 0.1                        & 0.04/3                     \\
    p$\overline 6m2$ & 0.15                 & 0.1/6                      & 0.08                       \\
    p$3m1$           & 0.15/2               & 0.1                        & 0.04                       \\ \hline
    \end{tabular}
    \caption{Tight-binding parameters of the SOC and the symmetry breaking model Hamiltonian terms in units of $t$. For each phase, the layer group with highest symmetry is given (See also Tab. \ref{tab:phase_overview}).}
    \label{tab:tb_param_LG}
\end{table}

%% file: supplement/DFT_supp.tex
\subsection{DFT Methods}
For our theoretical study of B, Al and Ga on SiC(0001) we employed state-of-the-art first-principles calculations based on density functional theory as implemented in VASP \cite{VASP1} within the PAW method \cite{VASP2,PAW}. For the exchange-correlation potential the PBE functional was used \cite{PBE} by expanding the Kohn-Sham wave functions into plane-waves up to an energy cut-off of 500 eV and 300 eV for the bulk calculations and for finite-size calculations, respectively. For the bulk calculations, we sampled the Brillouin zone on a $12 \times 12 \times 1$ regular mesh and SOC was self-consistently included \cite{SOC_VASP}. We consider a $(1\times 1)$ reconstruction of a triangular adatom monolayer adsorbed on the T1 position of Si-terminated SiC(0001) with an in-plane lattice constant of $3.07\,\text{\AA}$. The equilibrium structure is obtained by relaxing all atoms until all forces converged below $0.001\,\text{eV/\AA}$. For the bulk calculations, we consider four layers of SiC. To computationally access large lateral finite size systems, the substrate thickness is reduced to one layer of SiC. Electronic states arising from opposite surfaces are disentangled by a vacuum distance of at least 10 $\text{\AA}$ between periodic replicas in the $z$-direction. The dangling bonds of the substrate terminated surface are saturated with hydrogen atoms.

\subsection{DFT: Al on SiC Edge States}
\begin{figure}
    \centering
    \includegraphics[width=\columnwidth]{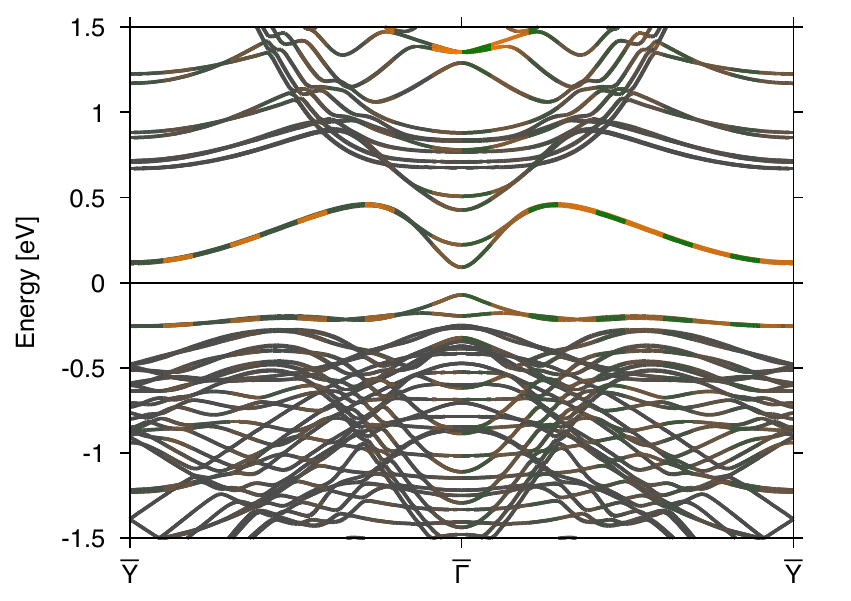}
    \caption{Band structure for the zigzag ribbon geometry. The color code denotes the edge character, shown for alternating edges (dashed lines).}
    \label{fig:supp_DFT_ribbon}
\end{figure}

Figure~\ref{fig:supp_DFT_ribbon} shows the band structure of a slab geometry with a zig-zag edge termination. The width of 12 unit cells is chosen to be comparable to the height of the triangular flake in Fig.~\ref{fig:DFT}. In agreement with the tight-binding model, the band structure is insulating and the edge states arising from opposite edges are energetically degenerate.

\subsection{Bulk Band Structures of B, Al and Ga on SiC}

\begin{figure*}
    \centering
    \includegraphics[width=\textwidth]{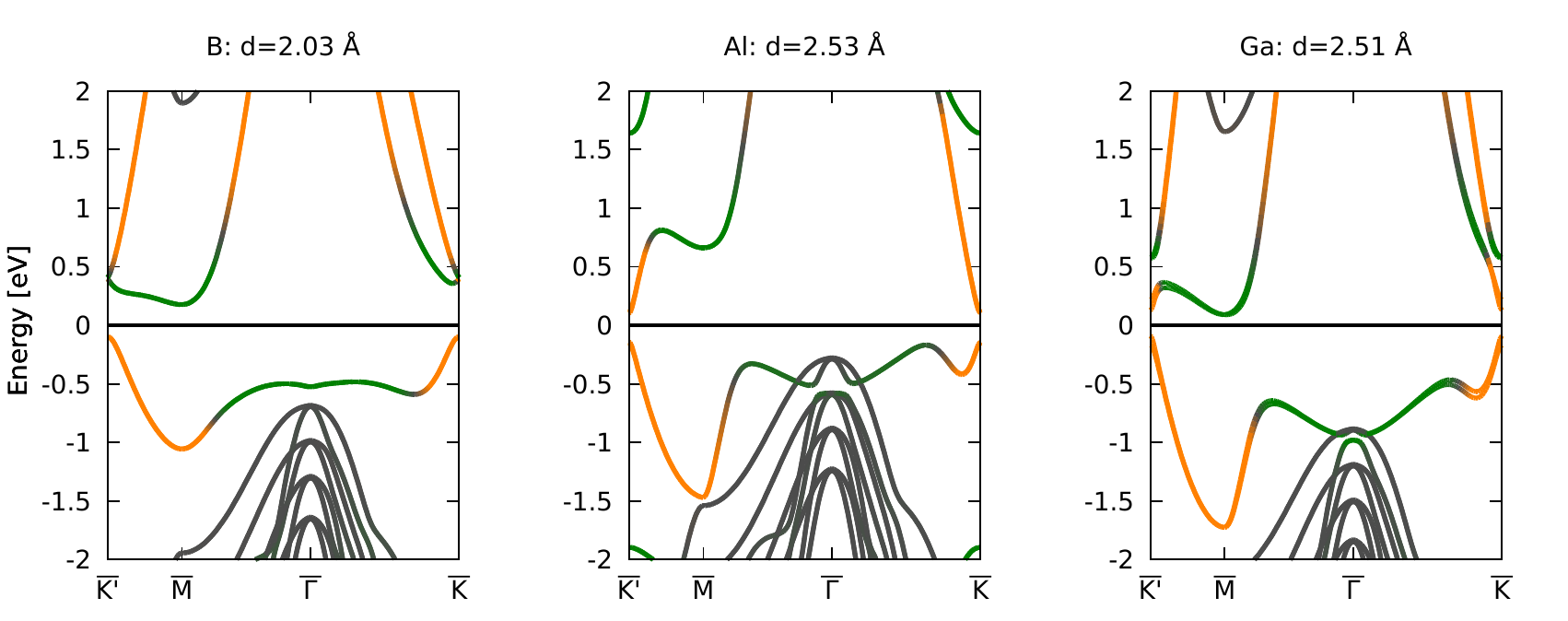}
    \caption{Orbital resolved band character and equilibrium distances of the three adsorbate systems on four layers of SiC (0001). The color code denotes the $p_{\pm}$ (orange) and $sp_z$ (green) orbital character.}
    \label{fig:supp_HOTI_bulk}
\end{figure*}

The orbital character projected bulk band structures of B, Al and Ga on SiC(0001) are shown in Fig.~\ref{fig:supp_HOTI_bulk}. All adsorbate systems show perfect qualitative agreement with the proposed HOTI model as they posses an insulating bulk band structure with massive in-plane Dirac cones at the valley momenta. The irreps of the valence bands are given in Table~\ref{tab:phase_overview_supp} for LG p$3m1$ and indicate a non-vanishing bulk dipole and quadrupole moment. The weak SOC interaction in B and Al results in almost two-fold degenerate bands, while the bands of the Ga monolayer possess a weak spin-splitting.

%% file: supplement/char_tab.tex
\subsection{Vertical Reflection Symmetry Breaking in Real and Reciprocal Space}
\label{sec:sup:vertical_reflection_sym}

\begin{figure*}[h]
        \centering
        \begin{subfigure}[b]{0.30\textwidth}
                \centering
                \includegraphics[width=\textwidth]{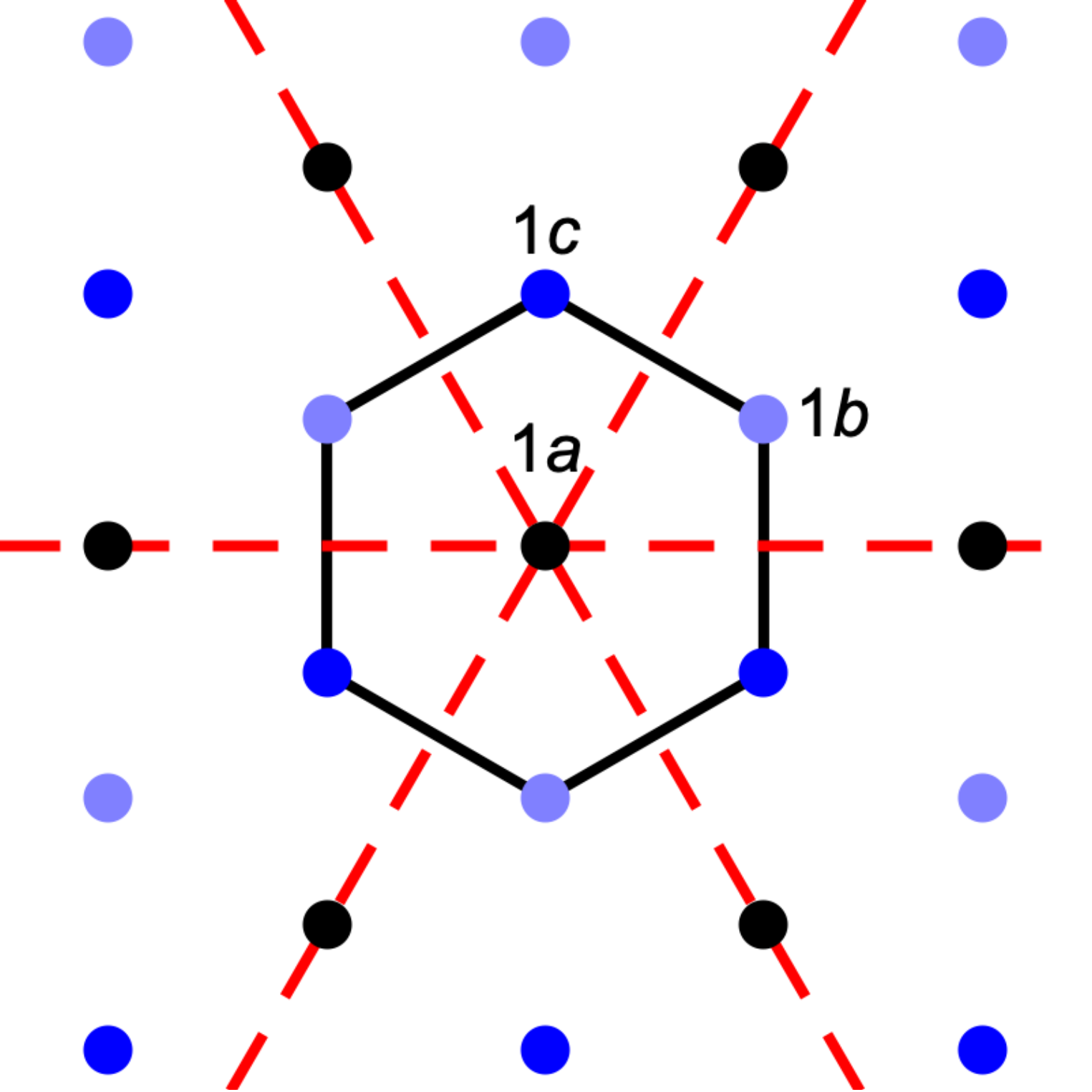}
                \caption{}
        \end{subfigure}
        \hfill
        \begin{subfigure}[b]{0.30\textwidth}
                \centering
                \includegraphics[width=\textwidth]{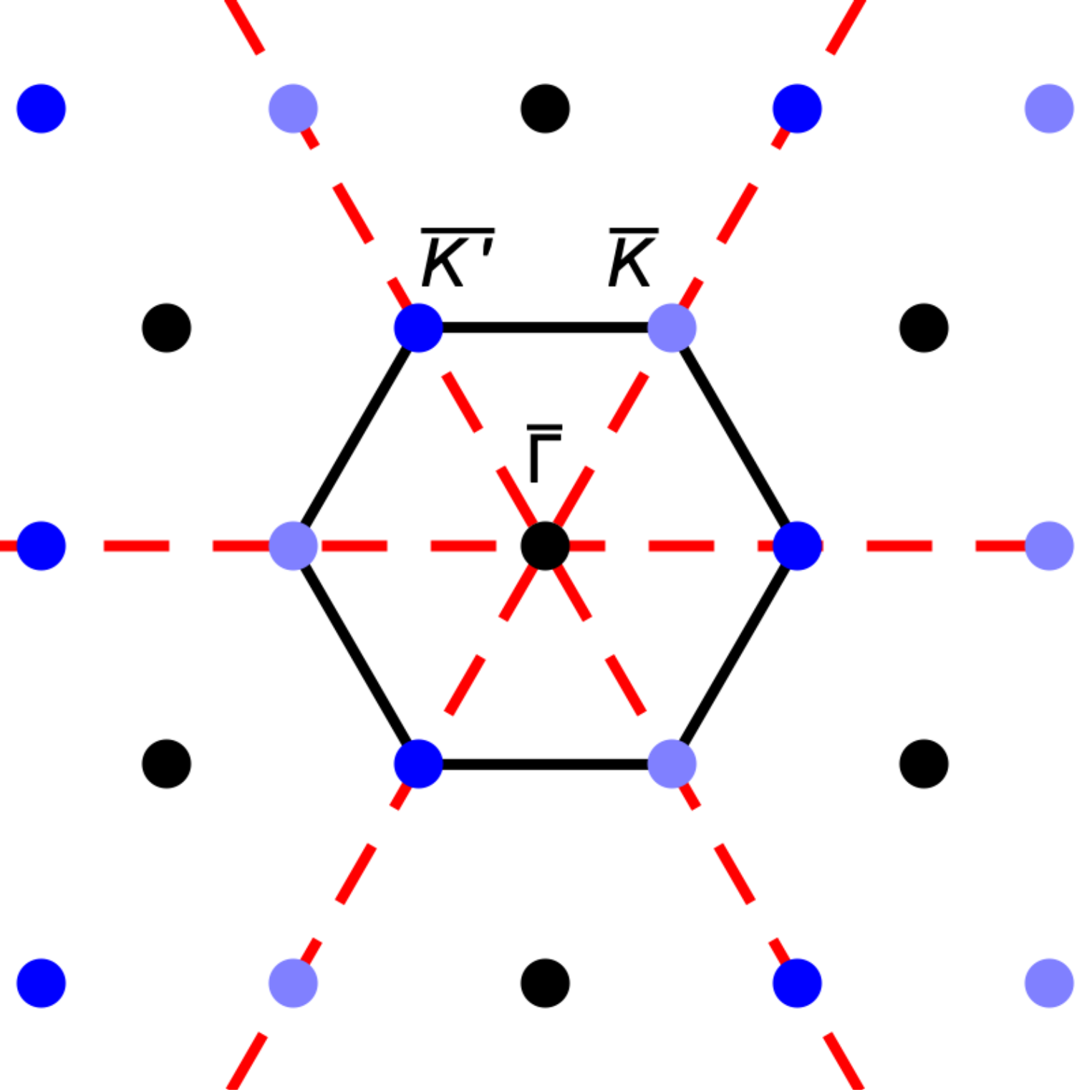}
                \caption{}
        \end{subfigure}
        \hfill
        \begin{subfigure}[b]{0.30\textwidth}
            \centering
            \includegraphics[width=\textwidth]{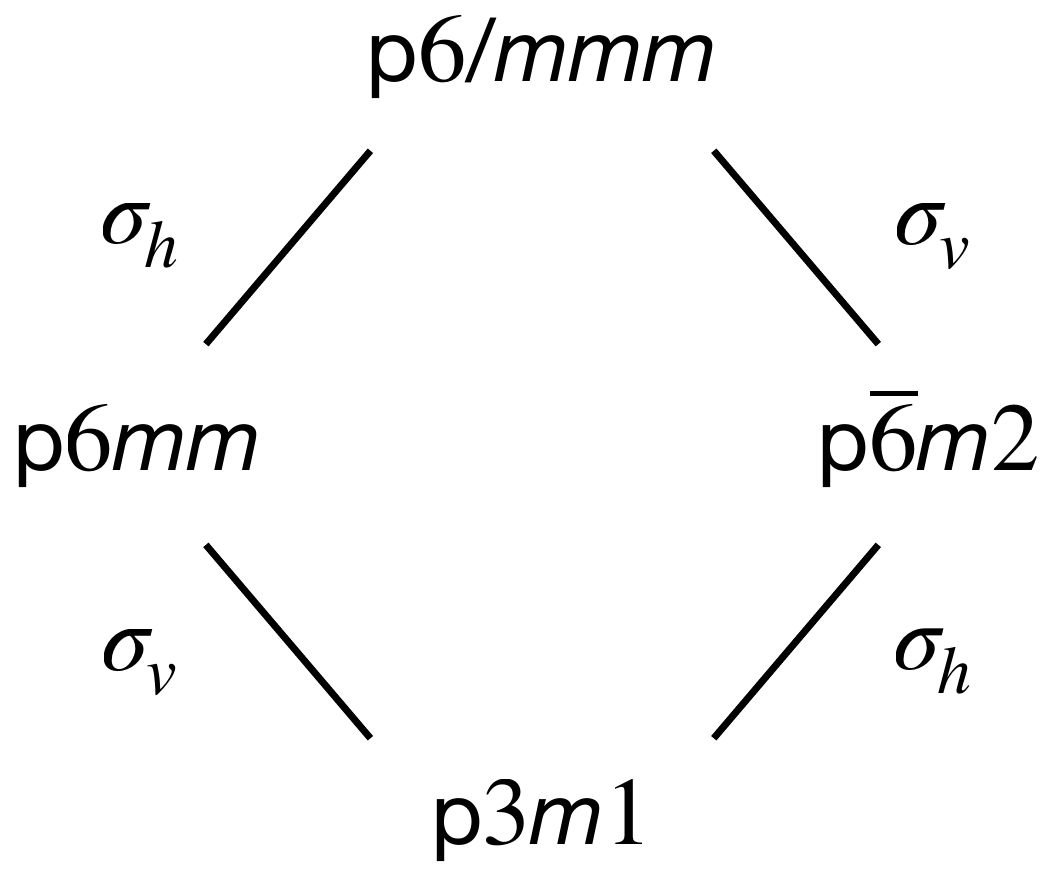}
            \caption{}
        \end{subfigure}
        \caption{(a,b) Impact of vertical reflection symmetry (red lines) on the hexagonal lattice. (a) The presence of $\sigma_v$ maps the Wyckoff position $1b$ onto $1c$; thus, in groups with $\sigma_v$, the two Wyckoff positions merge into a single position with multiplicity two. (b) The rotation of real space and reciprocal space lattice by $\pi/6$ against each other translates $\sigma_v$ into $\sigma_d$ in momentum space (see also Tab.~\ref{tab:littlegroup}). This introduces the vertical reflection to the little group of $1b$ and $1c$.
        (c) Real space layer group subgroup relation.}
        \label{fig:supp_hex_sym}
\end{figure*}

We now describe the role of the vertical reflection symmetry in real and reciprocal space. As illustrated in Fig.~\ref{fig:supp_hex_sym}, if vertical reflections (red lines in Fig.~\ref{fig:supp_hex_sym}a) are introduced, the LG p$3m1$ (or p$\overline{6}m2$) is promoted to p$6mm$ (or p$6/mmm$) (the relationship between layer groups is shown in Fig.~\ref{fig:supp_hex_sym}c). 
The vertical reflection planes map the Wyckoff position $1b$ onto $1c$ (notation refers to LG p$3m1$), which results in a single Wyckoff position with a multiplicity of two in the more symmetric group. 
This explains why the bulk dipole moment, which requires an asymmetric charge distribution with respect to the $1b$ and $1c$ Wyckoff positions, is only allowed when $\sigma_v$ is broken.

As the hexagonal real and reciprocal lattices are rotated relative to each other by $\pi/6$, the presence of $\sigma_v$ in real space translates into $\sigma_d$ in reciprocal space, as shown in Fig.~\ref{fig:supp_hex_sym}b. 
Thus, the reflection planes of $\sigma_d$ leave the $\overline{K}$ and $\overline{K}^\prime$ points invariant, which enlarges the little group of $\overline{K}$ and $\overline{K}^\prime$ from $3$ to $3m$.
Since the group $3$ only has one-dimensional single-valued irreps, in the absence of SOC, the $\sigma_v$-breaking term gaps the Dirac cones at $\overline{K}$ and $\overline{K}^\prime$.
In the presence of SOC, it can still drive a band inversion at the valley momenta which is identified by the irreps.
The little groups at each high symmetry point for each layer group are listed in Table~\ref{tab:littlegroup}.

%Focusing on the real space charge distribution and the wave function degeneracy at the valley momenta, an asymmetric charge distribution w.r.t. $1b$ and $1c$ and hence a bulk dipole moment can only be stabilized in the absence of $\sigma_v$. In reciprocal space, this is reflected by a lifted degeneracy of the former Dirac bands (without SOC) at $\overline{K}$ and $\overline{K}^\prime$ and the rotation eigenvalues dictate the localization of the elementary band representations for $\nu=0$.

\subsection{\label{sec:sup:character} Irreducible Band Representations}

Table~\ref{tab:phase_overview_supp} shows the irreps at high symmetry points for the topological phases shown in Fig.~\ref{fig:fig_Z2_bands}.
The labels of the irreps depend on the LG and can be derived from the characters tables shown in Tables~\ref{tab:littlegroup1}, \ref{tab:littlegroup2}, \ref{tab:littlegroup3}, \ref{tab:littlegroup4}, \ref{tab:littlegroup5} and \ref{tab:littlegroup6}. The notation follows Ref.~\cite{elcoro2017double}.
All of the band structures (B, Al and Ga on SiC) shown in Fig.~\ref{fig:supp_HOTI_bulk} are classified by the irreps of the HOTI phase in LG p$3m1$. 
\renewcommand{\arraystretch}{1.2}
\setlength{\tabcolsep}{2pt}
\begin{table*}
\centering
\begin{tabular}{ccccccccc}
\hline
Phase                               & Layer Group       & $\nu$ & SOC vs $\fsl{\sigma}_h$                       & SOC vs $\fsl{\sigma}_v$                       & IRREPs $\overline{\Gamma}$  & IRREPs $\overline{K}$                         & $\mathbf{P}=(P_1,P_2)$             & $Q_{12}$               \\ \hline
SOC insulator                       & p6/$mmm$          & 0     & $\lambda_\text{SOC}\gg\lambda_{\fsl{\sigma}_h}$ & $\lambda_\text{SOC}\gg\lambda_{\fsl{\sigma}_v}$ & $\overline{\Gamma}_{12}(2)$ & $\overline{K}_8(2)$                           & $(0,0)~\text{mod}~2$               & $0~\text{mod}~1$       \\ 
indenene-like $\fsl{\sigma}_h$ QSHI & p6$mm$            & 1     & $\lambda_\text{SOC}\ll\lambda_{\fsl{\sigma}_h}$ & $\lambda_\text{SOC}\gg\lambda_{\fsl{\sigma}_v}$ & $\overline{\Gamma}_9(2)$    & $\overline{K}_6(2)$                           & -                                  & -                      \\
$\fsl{\sigma}_v$ QSHI                    & p$\overline{6}m2$ & 1     & $\lambda_\text{SOC}\gg\lambda_{\fsl{\sigma}_h}$ & $\lambda_\text{SOC}\ll\lambda_{\fsl{\sigma}_v}$ & $\overline{\Gamma}_8(2)$    & $\overline{K}_7(1)\oplus\overline{K}_{12}(1)$ & -                                  & -                      \\ 
Triangular HOTI                     & p$3m1$            & 0     & $\lambda_\text{SOC}\ll\lambda_{\fsl{\sigma}_h}$ & $\lambda_\text{SOC}\ll\lambda_{\fsl{\sigma}_v}$ & $\overline{\Gamma}_6(2)$    & $\overline{K}_4(1)\oplus\overline{K}_6(1)$    & $(-\frac23,-\frac23)~\text{mod}~2$ & $\frac23~\text{mod}~1$ \\ \hline
\end{tabular}
\caption{Irreps and dipole/quadrupole moments of the insulating phases of Eq.~(\ref{eq:H_lattice}). For each phase, the layer group indicated is the highest symmetry group that satisfies the inequalities in columns four and five. The electric multipole moments in the $\nu=1$ phases are ill defined.
}
\label{tab:phase_overview_supp}
\end{table*}

%\subsection{ Character tables} JC: I absorbed this section into the above section on irreps
%In Table~\ref{tab:phase_overview}, the $C_3$ eigenvalues for the four phases and the corresponding layer groups of our model are summarized. 
%In Table~\ref{tab:phase_overview_supp}, we further show their symmetry irreps at $\overline\Gamma$ and $\overline K$; the little groups of these points are shown in Table~\ref{tab:littlegroup}. 
%The irreducible double-valued representations (i.e., both spinless and spinful representations) of these point groups are listed in the character tables in Tables~\ref{tab:littlegroup1}, \ref{tab:littlegroup2}, \ref{tab:littlegroup3}, \ref{tab:littlegroup4}, \ref{tab:littlegroup5} and \ref{tab:littlegroup6}. The notation follows Ref.~\cite{elcoro2017double}.

\begin{table*}[]
 \centering
 \begin{tabular}{*{5}{c}}
 \hline
 Real space layer group & $p6/mmm$ &$p6mm$ &$p\bar{6}2m$ & $p3m1$ \\
 
 Reciprocal space layer group & $p6/mmm$ &$p6mm$ &$p\bar{6}m2$ & $p31m$ \\
 %\hline 
 %Generators &$C_6$, $I$, $C_{2x}$ & $C_6$, $m_{100}$ & $C_6I$, $m_{100}$ & $C_3$, $m_{100}$\\
 \hline
 \hline
 Little group at $\overline\Gamma$ & $6/mmm$ & $6mm$ & $\bar{6}2m$ & $3m$ \\
 Little group at $\overline K$ &$\bar{6}m2$ & $3m$ & $\bar{6}$ & $3$ \\
 \hline
 \end{tabular}
\caption{Little groups at $\overline\Gamma$ and $\overline K$ for relevant layer groups.}
\label{tab:littlegroup}
\end{table*}

\begin{table*}[]
\centering
\scalebox{1.0}{
\begin{tabular}{c|*{18}{c}}
\hline
6/mmm& $1$ & $3_{001}$  & $2_{001}$ & $6_{001}$  & $2_{100}$ & $2_{1\bar1 0}$ & $d_1$ & $d_{3_{001}}$ & $d_{6_{001}}$  & $-1$ & $-3_{001}$ & $m_{001}$ & $-6_{001}$ & $m_{100}$ & $m_{1\bar 1 0}$ & $d_{-1}$ & $d_{-3_{001}}$ & $d_{-6_{001}}$ \\ \hline
$\Gamma_{1}^+$ & 1  & 1  & 1  & 1   & 1  & 1  & 1  & 1  & 1   & 1   & 1   & 1   & 1   & 1   & 1   & 1   & 1   & 1   \\
$\Gamma_{1}^-$ & 1  & 1  & 1  & 1   & 1  & 1  & 1  & 1  & 1   & -1  & -1  & -1  & -1  & -1  & -1  & -1  & -1  & -1  \\
$\Gamma_{2}^+$ & 1  & 1  & 1  & 1   & -1 & -1 & 1  & 1  & 1   & 1   & 1   & 1   & 1   & -1  & -1  & 1   & 1   & 1   \\
$\Gamma_{2}^-$ & 1  & 1  & 1  & 1   & -1 & -1 & 1  & 1  & 1   & -1  & -1  & -1  & -1  & 1   & 1   & -1  & -1  & -1  \\
$\Gamma_{3}^+$ & 1  & 1  & -1 & -1  & 1  & -1 & 1  & 1  & -1  & 1   & 1   & -1  & -1  & 1   & -1  & 1   & 1   & -1  \\
$\Gamma_{3}^-$ & 1  & 1  & -1 & -1  & 1  & -1 & 1  & 1  & -1  & -1  & -1  & 1   & 1   & -1  & 1   & -1  & -1  & 1   \\
$\Gamma_{4}^+$ & 1  & 1  & -1 & -1  & -1 & 1  & 1  & 1  & -1  & 1   & 1   & -1  & -1  & -1  & 1   & 1   & 1   & -1  \\
$\Gamma_{4}^-$ & 1  & 1  & -1 & -1  & -1 & 1  & 1  & 1  & -1  & -1  & -1  & 1   & 1   & 1   & -1  & -1  & -1  & 1   \\
$\Gamma_{5}^+$ & 2  & -1 & 2  & -1  & 0  & 0  & 2  & -1 & -1  & 2   & -1  & 2   & -1  & 0   & 0   & 2   & -1  & -1  \\
$\Gamma_{5}^-$ & 2  & -1 & 2  & -1  & 0  & 0  & 2  & -1 & -1  & -2  & 1   & -2  & 1   & 0   & 0   & -2  & 1   & 1   \\
$\Gamma_{6}^+$ & 2  & -1 & -2 & 1   & 0  & 0  & 2  & -1 & 1   & 2   & -1  & -2  & 1   & 0   & 0   & 2   & -1  & 1   \\
$\Gamma_{6}^-$ & 2  & -1 & -2 & 1   & 0  & 0  & 2  & -1 & 1   & -2  & 1   & 2   & -1  & 0   & 0   & -2  & 1   & -1  \\
$\overline\Gamma_{7}$  & 2  & -2 & 0  & 0   & 0  & 0  & -2 & 2  & 0   & 2   & -2  & 0   & 0   & 0   & 0   & -2  & 2   & 0   \\
$\overline\Gamma_{8}$  & 2  & 1  & 0  & $-\sqrt 3$ & 0  & 0  & -2 & -1 & $\sqrt 3$  & 2   & 1   & 0   & $-\sqrt 3$ & 0   & 0   & -2  & -1  & $\sqrt 3$  \\
$\overline\Gamma_{9}$  & 2  & 1  & 0  & $\sqrt 3$  & 0  & 0  & -2 & -1 & $-\sqrt 3$ & 2   & 1   & 0   & $\sqrt 3$  & 0   & 0   & -2  & -1  & $-\sqrt 3$ \\
$\overline\Gamma_{10}$ & 2  & -2 & 0  & 0   & 0  & 0  & -2 & 2  & 0   & -2  & 2   & 0   & 0   & 0   & 0   & 2   & -2  & 0   \\
$\overline\Gamma_{11}$ & 2  & 1  & 0  & $-\sqrt 3$ & 0  & 0  & -2 & -1 & $\sqrt 3$  & -2  & -1  & 0   & $\sqrt 3$  & 0   & 0   & 2   & 1   & $-\sqrt 3$ \\
$\overline\Gamma_{12}$ & 2  & 1  & 0  & $\sqrt 3$  & 0  & 0  & -2 & -1 & $-\sqrt 3$ & -2  & -1  & 0   & $-\sqrt 3$ & 0   & 0   & 2   & 1   & $\sqrt 3$ \\
\hline
\end{tabular}}
\caption{Character table for point group $6/mmm$.}
\label{tab:littlegroup1}
\end{table*}

\begin{table*}[]
\centering
\scalebox{1.0}{
\begin{tabular}{c|c c c c c c c c c }
\hline
$\bar 6 2m$& $1$ & $3_{001}$ & $m_{001}$ & $-6_{001}$ & $2_{100}$ & $m_{1\bar 10}$ & $d_1$ & $d_{3_{001}}$ & $d_{-6_{001}}$  \\ \hline
$\Gamma_1$ & 1  & 1  & 1  & 1   & 1  & 1  & 1  & 1  & 1   \\
$\Gamma_2$ & 1  & 1  & -1 & -1  & 1  & -1 & 1  & 1  & -1  \\
$\Gamma_3$ & 1  & 1  & -1 & -1  & -1 & 1  & 1  & 1  & -1  \\
$\Gamma_4$ & 1  & 1  & 1  & 1   & -1 & -1 & 1  & 1  & 1   \\
$\Gamma_5$ & 2  & -1 & 2  & -1  & 0  & 0  & 2  & -1 & -1  \\
$\Gamma_6$ & 2  & -1 & -2 & 1   & 0  & 0  & 2  & -1 & 1   \\
$\overline \Gamma_7$ & 2  & -2 & 0  & 0   & 0  & 0  & -2 & 2  & 0   \\
$\overline \Gamma_8$ & 2  & 1  & 0  & $-\sqrt{3}$ & 0  & 0  & -2 & -1 & $\sqrt{3}$  \\
$\overline \Gamma_9$ & 2  & 1  & 0  & $\sqrt{3}$  & 0  & 0  & -2 & -1 & $-\sqrt{3}$\\
\hline
\end{tabular}}
\caption{Character table for point group $\bar 6 2m$.}
\label{tab:littlegroup2}
\end{table*}

\begin{table*}[]
\centering
\scalebox{1.0}{
\begin{tabular}{c c c c c c c c c c }
\hline
6mm & $1$ & $3_{001}$  & $2_{001}$ & $6_{001}$ & $m_{100}$ & $m_{1\bar 1 0}$ & $d_1$ & $d_{3_{001}}$ & $d_{-6_{001}}$  \\ \hline
$\Gamma_1$ & 1  & 1  & 1  & 1   & 1  & 1  & 1  & 1  & 1   \\
$\Gamma_2$ & 1  & 1  & 1  & 1   & -1 & -1 & 1  & 1  & 1   \\
$\Gamma_3$ & 1  & 1  & -1 & -1  & -1 & 1  & 1  & 1  & -1  \\
$\Gamma_4$ & 1  & 1  & -1 & -1  & 1  & -1 & 1  & 1  & -1  \\
$\Gamma_5$ & 2  & -1 & 2  & -1  & 0  & 0  & 2  & -1 & -1  \\
$\Gamma_6$ & 2  & -1 & -2 & 1   & 0  & 0  & 2  & -1 & 1   \\
$\overline\Gamma_7$ & 2  & -2 & 0  & 0   & 0  & 0  & -2 & 2  & 0   \\
$\overline\Gamma_8$ & 2  & 1  & 0  & -$\sqrt{3}$ & 0  & 0  & -2 & -1 & $\sqrt{3}$  \\
$\overline\Gamma_9$ & 2  & 1  & 0  & $\sqrt{3}$  & 0  & 0  & -2 & -1 & -$\sqrt{3}$\\
\hline
\end{tabular}}
\caption{Character table for point group $6mm$.}
\label{tab:littlegroup3}
\end{table*}

\begin{table*}[]
\centering
\scalebox{1.0}{
\begin{tabular}{ c|c c c c c c c c c c c c }
\hline
$\bar 6$ & $1$ & $3_{001}^+$  & $3_{001}^-$  & $m_{001}$ & $-6_{001}^-$  & $-6_{001}^+$  & $d_1$ & $d_{3_{001}^+}$  & $d_{3_{001}^-
}$  & $d_{m_{001}}$ & $d_{-6_{001}^-}$ & $d_{-6_{001}^+}$ \\ \hline
$\Gamma_1$ & 1  & 1& 1& 1  & 1& 1& 1  & 1& 1& 1   & 1& 1\\
$\Gamma_2$ & 1  & 1& 1& -1 & -1  & -1  & 1  & 1& 1& -1  & -1  & -1  \\
$\Gamma_3$ & 1  & - $\bar\epsilon$ & - $\epsilon$ & 1  & - $\bar\epsilon$ & - $\epsilon$ & 1  & - $\bar\epsilon$ & - $\epsilon$ & 1   & - $\bar\epsilon$ & - $\epsilon$ \\
$\Gamma_4$ & 1  & - $\bar\epsilon$ & - $\epsilon$ & -1 &  $\bar\epsilon$  &  $\epsilon$  & 1  & - $\bar\epsilon$ & - $\epsilon$ & -1  &  $\bar\epsilon$  &  $\epsilon$  \\
$\Gamma_5$ & 1  & - $\epsilon$ & - $\bar\epsilon$ & 1  & - $\epsilon$ & - $\bar\epsilon$ & 1  & - $\epsilon$ & - $\bar\epsilon$ & 1   & - $\epsilon$ & - $\bar\epsilon$ \\
$\Gamma_6$ & 1  & - $\epsilon$ & - $\bar\epsilon$ & -1 &  $\epsilon$  &  $\bar\epsilon$  & 1  & - $\epsilon$ & - $\bar\epsilon$ & -1  &  $\epsilon$  &  $\bar\epsilon$  \\
$\overline\Gamma_7$ & 1  & -1  & -1  & -i & i& -i  & -1 & 1& 1& i   & -i  & i\\
$\overline\Gamma_8$ & 1  & -1  & -1  & i  & -i  & i& -1 & 1& 1& -i  & i& -i  \\
$\overline\Gamma_9$ & 1  &  $\bar\epsilon$  &  $\epsilon$  & -i & $\epsilon$  & $-\bar\epsilon$   & -1 & - $\bar\epsilon$ & - $\epsilon$ & i   & $-\epsilon$   & $\bar\epsilon$  \\
$\overline\Gamma_{10}$   & 1  &  $\bar\epsilon$  &  $\epsilon$  & i  & $-\epsilon$   & $\bar\epsilon$  & -1 & - $\bar\epsilon$ & - $\epsilon$ & -i  & $\epsilon$  & $-\bar\epsilon$   \\
$\overline\Gamma_{11}$   & 1  &  $\epsilon$  &  $\bar\epsilon$  & -i & $\bar\epsilon$  & $-\epsilon$   & -1 & - $\epsilon$ & - $\bar\epsilon$ & i   & $-\bar\epsilon$   & $\epsilon$  \\
$\overline\Gamma_{12}$   & 1  &  $\epsilon$  &  $\bar\epsilon$  & i  & $-\bar\epsilon$   & $\epsilon$  & -1 & - $\epsilon$ & - $\bar\epsilon$ & -i  & $\bar\epsilon$  & $-\epsilon$  \\
\hline
\end{tabular}}
\caption{Character table for point group $\bar 6$. $\epsilon=\frac{(1+i\sqrt 3)}{2}$.}
\label{tab:littlegroup5}
\end{table*}

\begin{table*}[]
\centering
\begin{tabular}{ c|c c c c c c }
\hline
3m & $1$ & $3_{001}$ & $m_{1\bar 1 0}$ & $d_1$ & $d_{3_{001}}$ & $d_{m_{1\bar 1 0}}$ \\ \hline
$\Gamma_1$ & 1  & 1  & 1  & 1  & 1  & 1  \\
$\Gamma_2$ & 1  & 1  & -1 & 1  & 1  & -1 \\
$\Gamma_3$ & 2  & -1 & 0  & 2  & -1 & 0  \\
$\overline\Gamma_4$ & 1  & -1 & -i & -1 & 1  & i  \\
$\overline\Gamma_5$ & 1  & -1 & i  & -1 & 1  & -i \\
$\overline\Gamma_6$ & 2  & 1  & 0  & -2 & -1 & 0 \\
\hline
\end{tabular}
\caption{Character table for point group $3m$.}
\label{tab:littlegroup4}
\end{table*}

\begin{table*}[]
\centering
\begin{tabular}{ c|c c c c c c c }
\hline
3 & $1$ &$3_{001}^+$  & $3_{001}^-$ &$d_1$ & $d_{3_{001}^+}$  & $d_{3_{001}^-}$  \\ \hline
$\Gamma_1$    & 1  & 1          & 1          & 1  & 1          & 1          \\
$\Gamma_2$    & 1  & -$\bar\epsilon$ & -$\epsilon$ & 1  & -$\bar\epsilon$ & -$\epsilon$ \\
$\Gamma_3$    & 1  & -$\epsilon$ & -$\bar\epsilon$ & 1  & -$\epsilon$ & -$\bar\epsilon$ \\
$\overline\Gamma_4$    & 1  & -1         & -1         & -1 & 1          & 1          \\
$\overline\Gamma_5$    & 1  & $\bar\epsilon$  & $\epsilon$  & -1 & -$\bar\epsilon$ & -$\epsilon$ \\
$\overline\Gamma_6$    & 1  & $\epsilon$  & $\bar\epsilon$  & -1 & -$\epsilon$ & -$\bar\epsilon$\\
\hline
\end{tabular}
\caption{Character table for point group $3$; $\epsilon=\frac{(1+i\sqrt 3)}{2}$.}
 \label{tab:littlegroup6}
\end{table*}